# Delta-doped β- Gallium Oxide Field Effect Transistor

Sriram Krishnamoorthy[1], Zhanbo Xia[1], Sanyam Bajaj[1], Mark Brenner[1], and Siddharth Rajan[1,2]

[1] *Electrical & Computer Engineering, The Ohio State University, Columbus, OH 43210, USA*

[2] *Materials Science and Engineering, The Ohio State University, Columbus, OH 43210, USA*

**Abstract:** We report silicon delta doping in Gallium Oxide (β-$Ga_2O_3$) grown by plasma assisted molecular beam epitaxy using a shutter pulsing technique. We describe growth procedures that can be used to realize high Si incorporation in an oxidizing oxygen plasma environment. Delta doping was used to realize thin (12 nm) low-resistance layers with sheet resistance of 320 Ohm/square (mobility of 83 $cm^2$/Vs, integrated sheet charge of $2.4 \times 10^{14}$ $cm^{-2}$). A single delta-doped sheet of carriers was employed as a channel to realize a field effect transistor with current $I_{D,MAX}$ =292 mA/mm and transconductance $g_M$ = 27 mS/mm.

a) Authors to whom correspondence should be addressed.
Electronic mail: krishnamoorthy.13@osu.edu, rajan@ece.osu.edu

β-$Ga_2O_3$ is an emerging ultra-wide band gap semiconductor, which holds promise for next generation wide band gap devices[1] due to the availability of high quality substrates grown by melt-based techniques[2,3], and the ability to achieve wide range of conductivity[4,5]. Epitaxial growth of β-$Ga_2O_3$ has been reported using growth techniques such as MBE[4,6], MOCVD[7], HVPE[8] and LPCVD[9]. The estimated breakdown field of 8 MV/cm[10] and the calculated low impact ionization coefficient[11] make $Ga_2O_3$ attractive for power switching applications with high blocking voltage capability. β-$Ga_2O_3$- based devices that have been reported in the literature consist of thick (~ 200-300 nm) low-doped channels realized using Si ion implantation[12,13,14], and doping by MBE [15] or MOCVD [16,17] epitaxial growth techniques. Mechanical exfoliation[18,19,20] and transfer has also been used to realize FETs based on β-$Ga_2O_3$.

In addition to the potential for high blocking voltage for power switching applications, β-$Ga_2O_3$ could also be used to realize power amplifiers for RF and mm-wave applications. The high breakdown field of β-$Ga_2O_3$ could potentially enable higher breakdown voltage and higher channel charge density than the state-of-art GaN-based HEMTs, and the availability of low dislocation density substrates can enable high reliability at high fields. However, challenges related to the poor thermal conductivity[21] of β-$Ga_2O_3$ (which is significantly lower than SiC substrates used in AlGaN/GaN HEMTs), and surface passivation will need to be overcome for such devices.

Since lateral high frequency FETs require good aspect ratio, high charge density channel layers are critical to evaluating the performance of β-$Ga_2O_3$ for RF applications. In this work, we describe techniques for epitaxial delta doping of high sheet charge density channel layers, and evaluate the transport and device performance of these layers. Well-designed growth protocols can enable silicon delta-doping in β-Ga$_2$O$_3$ grown using plasma-assisted MBE with high mobility and low sheet resistance.

We first evaluated the growth rate for our system using a combination of capacitance voltage and SIMS measurements. Homoepitaxial β-Ga$_2$O$_3$ was grown on (010) – oriented bulk substrate using O$_2$ plasma-assisted molecular beam epitaxy equipped with standard effusion cells for gallium, silicon and an RF plasma source for oxygen . The (010) orientation of β-Ga$_2$O$_3$ was chosen for the study as it has been shown to support a higher growth rate[4,22] than other planes such as (-201) and (100) β-Ga$_2$O$_3$. Growth was initiated on commercially obtained β-Ga$_2$O$_3$ templates[23] with the growth of undoped β-Ga$_2$O$_3$ at a substrate temperature of 700° C (measured by thermocouple), plasma power of 300W and chamber pressure of 1.5x10$^{-5}$ Torr. The growth rate was estimated by growing undoped films on Sn-doped (010) n-type substrates, and performing capacitance-voltage (C-V) measurements. Pt/Au/Ni Schottky contacts were evaporated on the grown films, and Ti/Au/Ni Ohmic contacts were evaporated on the back surface of the sample after a BCl$_3$/Ar- chemistry based inductively coupled plasma reactive ion etch (ICP-RIE ) process[10]. The thickness was estimated from the depletion width measured from the C-V profiles, which indicated a fully depleted epitaxial layer. Ga flux of 4x10$^{-8}$ Torr and 8x10$^{-8}$ Torr resulted in a Ga-limited growth rate of 155 nm/hour and 240 nm/hour, respectively. The film thickness/ growth rate extracted from the CV measurements matched with the thickness estimated using secondary ion mass spectrometry (SIMS). Smooth surface morphology with a rms roughness of 0.75 nm was measured using atomic force microscopy (Fig. 1).

To evaluate the effect of silicon oxidation in the plasma environment, we investigated the Si incorporation in the films as a function of time/thickness and Si cell temperature. Ga flux of 8x10$^{-8}$ Torr corresponding to a growth rate of 240 nm/hour was used. The Si cell temperature used for growth (Fig. 2 (a)) is shown together with the Si incorporation measured using SIMS (Fig. 2(b)). Between each layer, growth was interrupted, all shutters were closed, and the Si cell temperature was raised to 1290°C to desorb any oxide that may have formed on the surface of the solid Si source. The measured secondary ion mass spectrometry (SIMS) profile is shown in Fig. 2 (b). Although a high Si doping concentration (~ $6 \times 10^{18}$ cm$^{-3}$ – $10^{21}$ cm$^{-3}$) is seen at the start of each Si-doped layer, the doping concentration dropped rapidly as the growth proceeded. We attribute this to oxidation of the Si source leading to reduction of Si flux during the growth of each layer. Therefore, growth of thick Si-doped layers by MBE using a solid Si source is limited by source oxidation. A second limitation is related to the growth of the film itself. Continuous growth with Si cell temperatures above 850°C was found to result in dimming and disappearance of reflection high energy electron diffraction (RHEED) patterns, indicating a loss of crystallinity at high doping levels.

Therefore, to avoid source oxidation and to obtain flat doping profiles, a shutter pulsing scheme was adopted. The Si source was pulsed (1s with duty cycle of 1 minute) while maintaining the Si source at relatively higher temperature (850°C – 950° C), with the Ga and O shutters open throughout (Fig. 3 (a)). Such an approach is expected to result in delta doping of Si, with undoped Ga$_2$O$_3$ spacers. The SIMS profile of Si-doped Ga$_2$O$_3$ layers grown using the delta doping/pulsed doping approach is shown in Fig. 3 (b). Near-flat and abrupt Si doping profiles with average doping concentrations in the range $4 \times 10^{17}$ cm$^{-3}$ to $4 \times 10^{19}$ cm$^{-3}$ were observed in the SIMS measurement.

Electrical characteristics of the delta-doped layers were investigated using an epitaxial structure consisting of undoped β-Ga$_2$O$_3$ buffer layer and delta-doped layers grown on Fe-doped semi-insulating (010) oriented β-Ga$_2$O$_3$ substrates (Figure 4(a)). The growth conditions were similar to the doping experiments described earlier, with a growth rate of 4 nm/minute. Silicon delta-doped layers were separated by 4 nm undoped $Ga_2O_3$ layers. Energy band and charge profile diagrams of the delta-doped layer are shown in Figure 4(b) Since the thickness of the spacer layers is of the same order as that of the Debye length, Fermi level remains above the conduction band edge throughout the delta-doped region, with no barriers to out-of-plane (vertical) transport. Delta-doped layers can therefore be used as an efficient method to realize high conductivity access regions[24].

Ohmic contacts were patterned using stepper lithography and Ti/Au/Ni metal stack was deposited after $BCl_3$/Ar treatment for Ohmic contact formation. Device isolation was performed by ICP-RIE etch using a $BCl_3$-based chemistry. Ohmic contacts were annealed at 470° C for 1 minute. Contact resistance of 0.35 ohm mm with a low specific resistance of $4.3 \times 10^{-6}$ $\Omega$ $cm^2$ was measured using TLM structure (Figure 4(c)). Our results suggest that delta-doped layers can be used as excellent contact layers. Hall measurements indicated a total charge of $2.4 \times 10^{14}$ $cm^{-2}$ with a high mobility of 83 $cm^2$/Vs and a sheet resistance of 320 ohms/sq. Mobility in the range of 77 – 81 $cm^2$/Vs was measured in multiple samples grown using the pulsed-doping approach, with a measured sheet charge density ranging from $2.7 \times 10^{14}$ $cm^{-2}$ – $3.5 \times 10^{14}$ $cm^{-2}$ (volume charge density of $6.8 \times 10^{19}$ $cm^{-3}$ – $1.7 \times 10^{20}$ $cm^{-3}$). Such mobility values are higher than reported mobility values in the literature[4, 22,25] even for a relatively lower doping concentration (< 60 $cm^2$/Vs for a doping concentration of ~$3 \times 10^{19}$ $cm^{-3}$). The carrier mobility is expected to be higher than uniformly doped material with the same equivalent doping density due to the spread of the electron population into undoped regions, which leads to lower impurity scattering[26].

To evaluate the application of delta doping for field effect transistors, the device structure shown in Figure 5(a) was used. The top two delta-doped layers were etched ($BCl_3$-based ICP-

RIE) in the channel region to realize a channel with one delta-doped sheet, while all three sheets were left intact for the contact region. A gate dielectric (Al$_2$O$_3$) was deposited using atomic layer deposition (ALD), followed by patterning and evaporation of Pt/Au/Ni gate stack. The ALD process was similar to that described earlier[27]. Channel charge density of 3.3x10$^{13}$ cm$^{-2}$ with a mobility of 34 cm$^2$/ Vs was measured using hall measurements after the gate recess. The reduction in electron mobility after the gate recess could be due to etching-induced surface damage and remote scattering[28], but further investigation is required to confirm the origin of this effect.

Output characteristics and transfer characteristics (V$_{DS}$ = 7 V) of the delta-doped FET are shown in Figures 5(b) and 5(c). Current (charge) modulation and channel pinch-off were clearly observed in the output characteristics. A very high current (I$_{D,max}$ = 290 A/mm) was measured at V$_{DS}$ = 7 V and V$_G$ = +2 V. Channel pinch off was observed at a gate bias of -14 V (I$_D$ = 58 μA/mm at V$_{DS}$ = 7 V). A peak transconductance of 27 mS/mm was measured at a gate bias of -4.5 V. Flat capacitance voltage curves (not shown here) confirmed the spatial confinement of the charge (~ 22-24 nm from the gate metal) in the delta-doped layer. However, channel pinch – off could not be observed in C-V characteristics of large area C-V test structures due to gate leakage at large negative gate bias. The current and transconductance values observed in this device are much higher than previously reported epitaxial β-Ga$_2$O$_3$ FETs.

In summary, we report delta doping / pulsed doping as a promising approach to achieving high electron mobility (~ 80 cm$^2$/Vs) at high charge density (2.4 x 10$^{14}$ cm$^{-2}$) . Delta doping enables scaling of the gate-channel distance, which is critical for high frequency devices and can enable high conductivity contact layers for transistors and other device applications. We also report a delta-doped β-Ga$_2$O$_3$ with I$_{D,max}$ = 292 mA/mm and g$_{m,max}$ = 27 mS/mm. The results reported in this work indicate the promise of delta-doping based devices for high current density/mobility and could enable new device architectures for high performance $\beta$-$Ga_2O_3$-based high frequency devices.

**Acknowledgements**

We acknowledge funding from Office of Naval Research, Grant Number N00014-12-1-0976 (EXEDE MURI). We thank Air Force Research Laboratory, WPAFB, Dayton Ohio for support.

**Figure captions**

Figure 1: Atomic force microscopy image of MBE-grown (010) β-Ga$_2$O$_3$ surface with a rms roughness of 0.75 nm.

Figure 2: SIMS profile of Si dopants showing strong oxidation of the source and reduction of Si flux during the growth. Source was heated up to 1290$^\circ$C in between the growth of each layers to remove oxide from source (Marked as red).

Figure 3: (a) Shutter pulsing scheme used to avoid Si source oxidation during growth (b) SIMS profile of Si dopants showing near-flat profile while using the delta doping approach.

Figure 4: (a) Epitaxial structure, (b) equilibrium band diagram and charge profile, and (c) TLM measurements on delta-doped FET structure.

Figure 5: (a) Device schematic, (b) Output characteristics, and (c) transfer characteristics of delta-doped FET device.

**Figures**

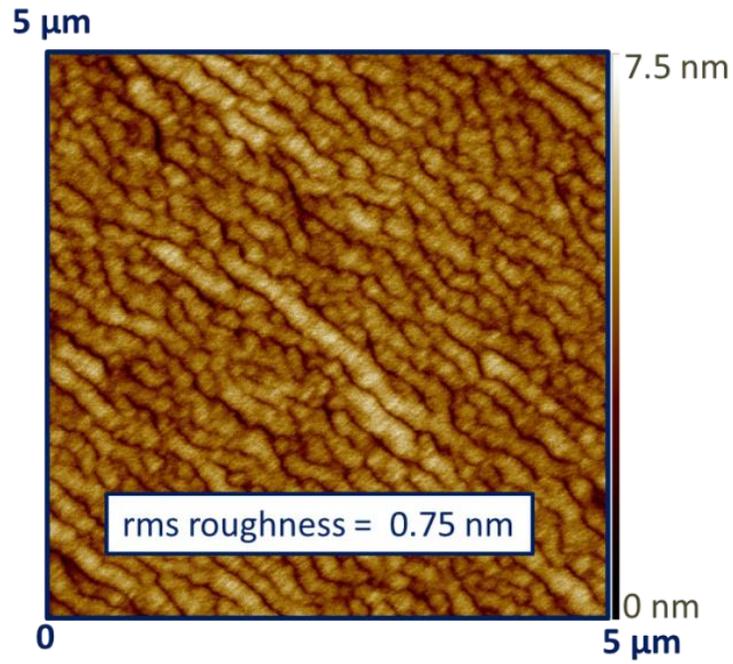

**Figure 1**

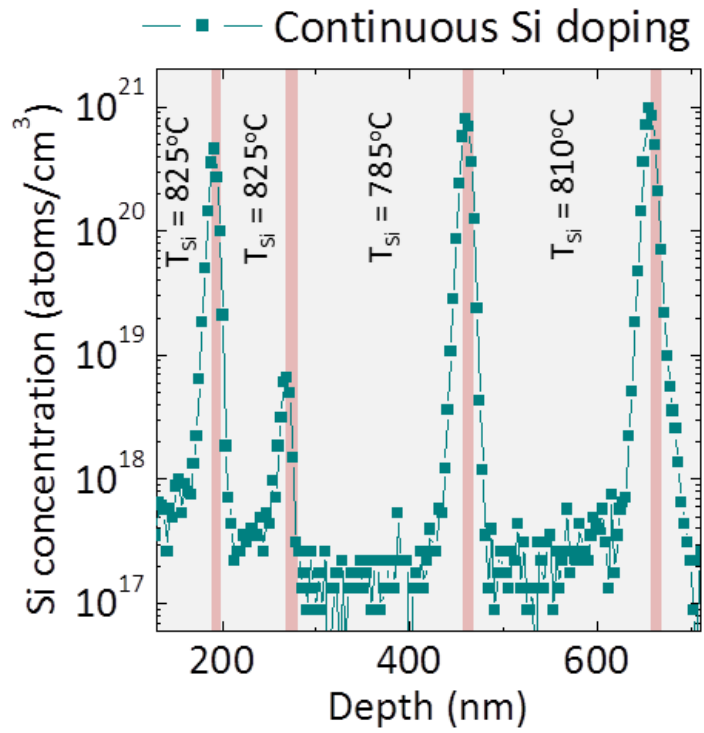

**Figure 2**

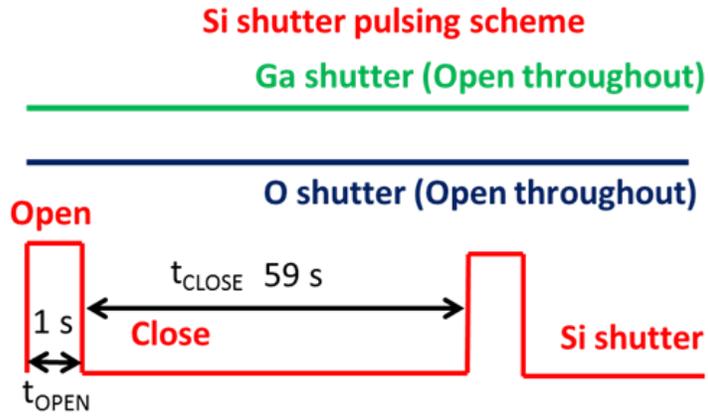

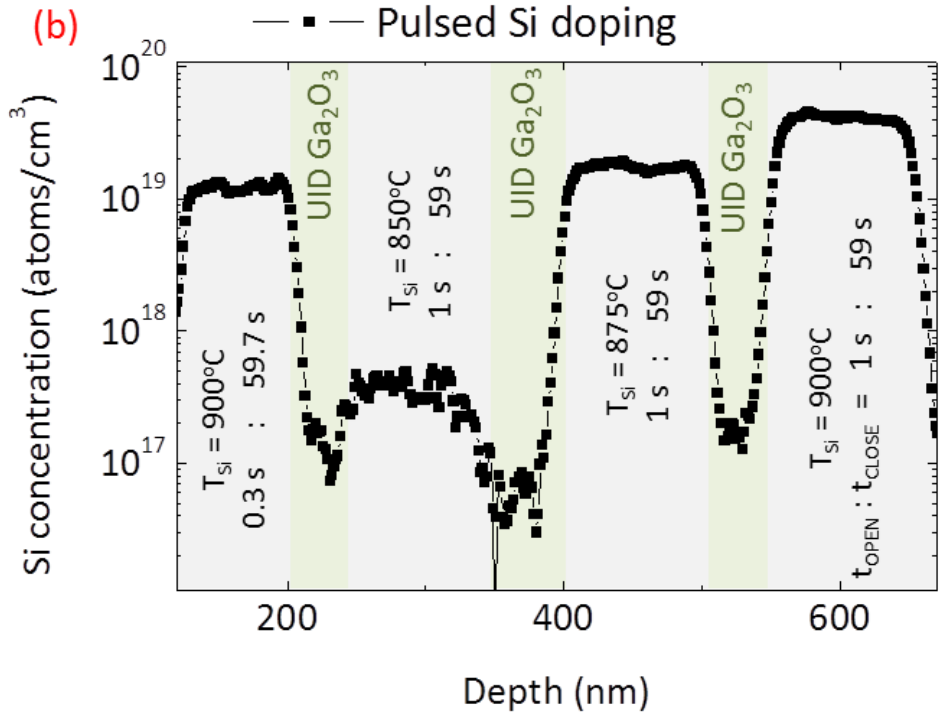

**Figure 3**

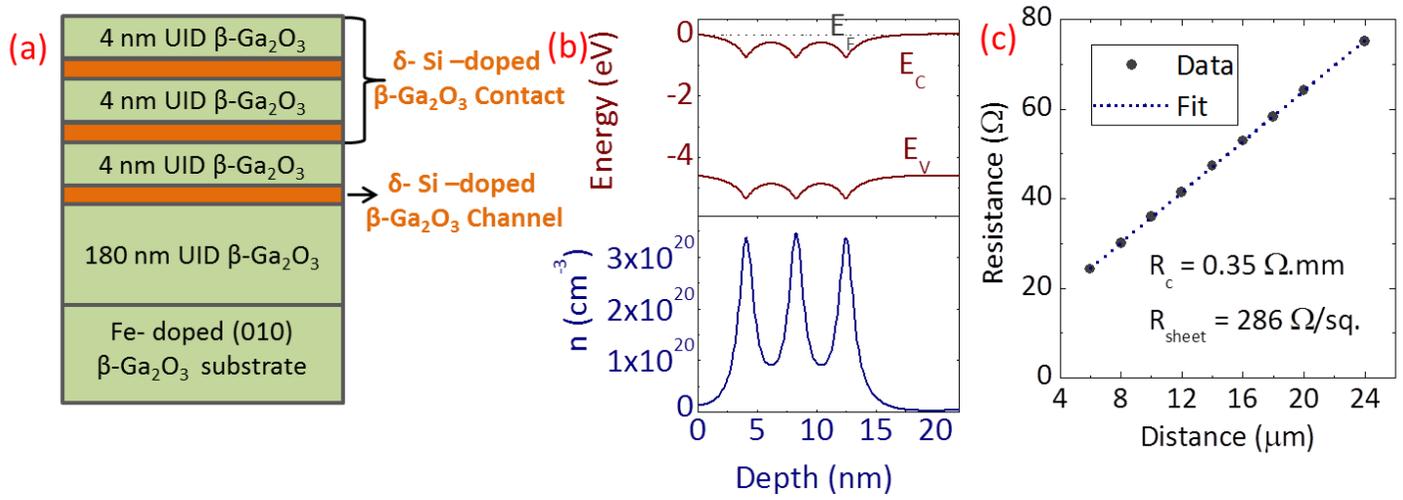

Figure 4.

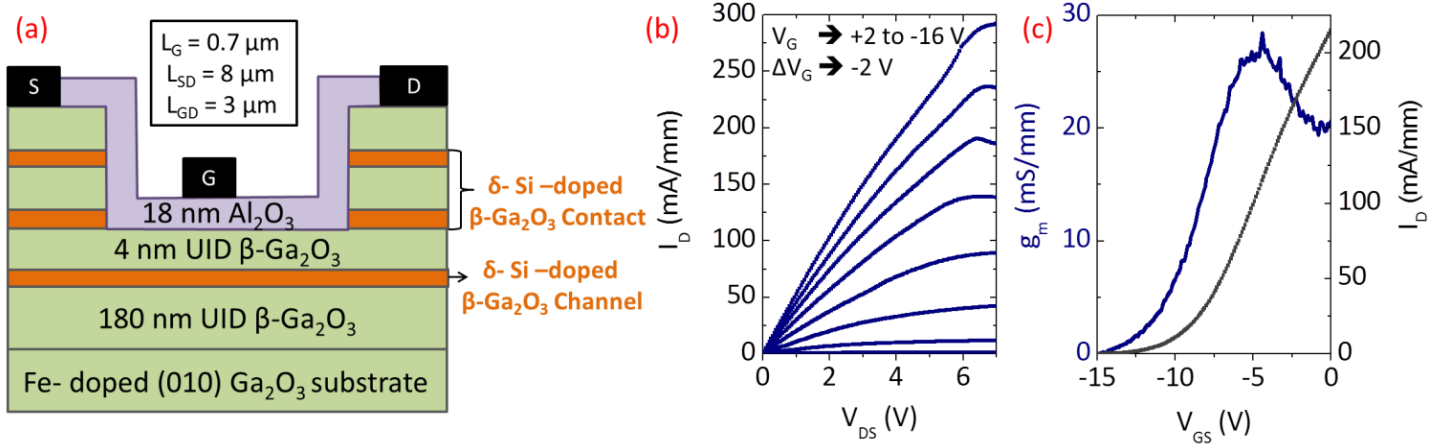

Figure 5.